# Generalized functions as a tool for nonsmooth nonlinear problems in mathematics and physics

J.F.Colombeau,( jf.colombeau@wanadoo.fr).

Abstract: A differential algebra of nonlinear generalized functions is presented as a tool for a wide range of nonsmooth nonlinear problems. The power of the differential algebra is used to do mathematical calculations or proofs; then the final result is often a classical function or distribution which is not solution in the classical or distributional sense. The aim of this text is to invite the listeners in applying this tool in their own research without significant prerequisites by presenting its use on a sample of elementary applications from mathematics and physics. This text is the written version of a talk at the 2007 annual meeting of the AMS, NewOrleans, january 2007.

Math. Subject Classification. primary: 35D05, secondary: 35D10,35L67,35Q72,46F30, 35R05, 65M99, 73J30, 82D20, 83C57.

Key words: algebras of generalized functions, nonsmooth nonlinear problems, shock waves, heat equation, ill-posed problems.

**1-Prerequisites**. Let $\Omega$ be an open set in $\mathbb{R}^N$. "Distributions" are a generalization of the concept of function [Ho]. If $\mathbf{D'}(\Omega)$ denotes the vector space of all distributions on $\Omega$ and $\mathbf{L}^p_{loc}(\Omega)$ the vector space of functions f such that $\int_K |f(x)|^p \, dx < +\infty$ for every compact set K in $\Omega$ ($1 \leq p < +\infty$; the case $p=+\infty$ works also with, as usual, the sup. norm), one has $\mathbf{L}^p_{loc}(\Omega) \subset \mathbf{D'}(\Omega)$. We will not need to enter any more into the distributions but simply note that the elements of $\mathbf{D'}(\Omega)$ have many of the properties of the $\mathbf{C}^\infty$ functions on $\Omega$ concerning differentiation: one defines internal partial derivatives: $\frac{\partial}{\partial x_i}\mathbf{D'}(\Omega) \subset \mathbf{D'}(\Omega)$.

In short: the concept of distributions allows one to freely differentiate rather irregular functions (that are not differentiable in the classical sense) at the price that their partial derivatives are objects (distributions) that are not usual functions. The typical example is given by the Heaviside function H. This is defined by: H(x) = 0 if x < 0, H(x) = 1 if x > 0 (and H(0) unspecified). H is not differentiable at x = 0 in the classical sense because of the discontinuity there. However its derivative, in the sense of distributions, is the "Dirac delta function" $\delta$: $\delta(x)=0$ if $x \neq 0$, $\delta(0)$ "infinite" so that
$$\int \delta(x)dx = [H]_{-\infty}^{+\infty} = 1-0 = 1.$$
The above explains calculations in which physicists differentiate functions that, like H, cannot be differentiated in the classical sense. Physicists not only differentiate irregular functions, they also mix differentiation and multiplication by treating formally irregular functions as if they were $\mathbf{C}^\infty$ functions. This motivates the introduction of nonlinear generalized functions [Bia,C1,C2,C3,C4,G-K-O-S,N-P-S,O4,S-V].This theory can be understood without prerequisites since its simplest formulation requires only elementary calculus [C3 p163,C4 p261,G-K-O-S p10,S-V p99]. The nonlinear generalized functions form a differential algebra $\mathbf{G}(\Omega)$ (i.e. an algebra with internal partial derivatives: $\frac{\partial}{\partial x_i} \mathbf{G}(\Omega) \subset \mathbf{G}(\Omega)$, in the situation $\mathbf{C}^\infty(\Omega) \subset \mathbf{D'}(\Omega) \subset \mathbf{G}(\Omega)$ in which the inclusion $\mathbf{D'}(\Omega) \subset \mathbf{G}(\Omega)$ is canonical (i.e. free from arbitrary choices) and

- the partial derivatives $\frac{\partial}{\partial x_i}$ in $\mathbf{G}(\Omega)$ induce those in $\mathbf{D'}(\Omega)$
- the multiplication in $\mathbf{G}(\Omega)$ induces on $\mathbf{C}^\infty(\Omega)$ the usual multiplication.

The objects in $\mathbf{G}(\Omega)$ can be treated as $\mathbf{C}^\infty$ functions on $\Omega$, but not always exactly like $\mathbf{C}^\infty$ functions, which explains various inconsistencies encountered by physicists from "formal" calculations. The aim of the talk is to show how these nonlinear generalized functions can be used as a tool for nonsmooth nonlinear problems and for ill-posed inverse problems.

The need of a novelty had been perceived by Schwartz in his "impossibility result" [C1] stating the definitive impossibility to multiply distributions in any mathematical context. We interpret this need of novelty by introducing a "weak equality" or "association" such that two associated elements of $\mathbf{G}(\Omega)$ are not necessarily equal: in $\mathbf{G}(\Omega)$ we say that $G_1 \approx G_2$ ("$G_1$ associated to $G_2$") if $\forall \varphi \in \mathbf{C}_c^\infty(\Omega)$ (i.e. infinitely differentiable with compact support) $\int_\Omega (G_1-G_2)(x)\varphi(x)dx$ is "infinitesimal" (i.e. a concept of number such that its absolute value is less than any strictly positive real number, and that may be nonzero). Schwartz impossibility result stems from the fact he demanded (in our language) that $G_1 \approx G_2$ implies $G_1 = G_2$ (for some particular $G_1, G_2$, see[C1]). Not only this is not useful but also physics displays instances in which several different Heaviside functions (associated to each other) are requested to model different physical variables: see the example of elastoplastic shock waves [Bia p120, C3 p106, C5, C-LR]. Thus, in mathematics capable to model physics, $G_1 \approx G_2$ should not imply $G_1 = G_2$ (even for very simple $G_1, G_2$: Heaviside functions).

Let us also show this fact mathematically: compute the integral
$$I = \int (H^2(x) - H(x)) \cdot H'(x) dx$$
where H denotes the Heaviside step function and H' its derivative (the Dirac delta distribution). H may be considered as an idealization of a $\mathbf{C}^1$ function with a jump from the value 0 to the value 1 in a tiny interval around $x = 0$. Thus classical calculations are justified: $I = [H^3/3 - H^2/2]_{-\infty}^{+\infty} = 1/3 - 1/2 = -1/6$. This implies that $H^2 \neq H$ (since $I \neq 0$): $H^2$ and $H$ differ at $x = 0$, precisely where $H'$ takes an "infinite value", and this undefined form $0 \times \infty$ gives here the value -1/6 after integration. Therefore the classical formula $H^2 = H$ has to be considered as erroneous in a context suitable to compute I. But it holds in the sense that $\forall \varphi \in \mathbf{C}_c^\infty(\Omega)$ $\int (H^2(x) - H(x)) \cdot \varphi(x) dx$ is infinitesimal, i.e. $H^2 \approx H$ (notice that $H^2 \neq H$). This means that the = needed should be more refined than the classical concept of equality which is replaced by $\approx$.

For physical applications it will be basic to have in mind that although there is only one Heaviside distribution, there is an infinity of Heaviside like objects in $\mathbf{G}$ (in particular any $H^N$); all of them are called "Heaviside generalized functions"; of course the same holds for their derivatives: "Dirac delta generalized functions". The $\mathbf{G}$-context retains an "internal structure" for these objects at the same time as it provides an idealization needed for the sake of simplicity. The fact these objects have no internal structure in $\mathbf{D'}(\Omega)$ and the existence of a canonical inclusion $\mathbf{D'}(\Omega) \subset \mathbf{G}(\Omega)$ looks paradoxical: it comes from the fact that an object in $\mathbf{D'}(\Omega)$ is, concerning this inclusion, considered with "all possible internal structures" at the same time (this makes sense from the definitions and finally amounts to an absence of internal structure).

Therefore if we denote by $\mathbf{D'}(\Omega)^\approx$ the vector space of those elements of $\mathbf{G}(\Omega)$ associated to a distribution one has the set of inclusions:

$$\mathbf{C}^\infty(\Omega) \subset \mathbf{D'}(\Omega) \subset \mathbf{D'}(\Omega)^{\approx} \subset \mathbf{G}(\Omega).$$

The partial derivatives $\dfrac{\partial}{\partial x_i}$ are internal in all these four spaces, but only $\mathbf{C}^\infty(\Omega)$ and $\mathbf{G}(\Omega)$ are algebras. If N>1 $H^N$ is in $\mathbf{D'}(\Omega)^{\approx}$ and not in $\mathbf{D'}(\Omega)$, $\delta^N$ is in $\mathbf{G}(\Omega)$ and not in $\mathbf{D'}(\Omega)^{\approx}$.

See [C1] as another recent introductory talk to these generalized functions for a different audience. In the literature one finds mainly two types of "**G**-algebras": a "simplified" or "special" algebra $\mathbf{G}_S(\Omega)$ (the inclusion of $\mathbf{D'}(\Omega)$ into $\mathbf{G}_S(\Omega)$ is not canonical) and the "full" algebra $\mathbf{G}(\Omega)$; see [S-V p111] for the respective roles of $\mathbf{G}_S(\Omega)$ and $\mathbf{G}(\Omega)$ in physics.

**2- Short examples that stress the roles of = and $\approx$.**
**\*A scalar conservation law.** Consider the equation classically stated as
    (1)    $u_t + u.u_x = 0$,
(indices mean t and x derivatives respectively) and seek a solution in form of two constant states $u_l$ and $u_r$ separated by a discontinuity with constant speed c. This solution is represented by the formula
    (2)    $u(x,t) = u_l + (u_r - u_l)H(x-ct)$,
where H is any Heaviside generalized function. Putting formula (2) into equation (1) one finds at once $c = (u_r + u_l)/2$. Now multiplication of (1) by u gives:
    (3)    $u.u_t + u^2.u_x = 0$.
Putting (2) into (3) one finds a different value of c. Where is the mistake? The above proves that (2) is only solution of the weaker statement
$$u_t + u.u_x \approx 0 \text{ of (1)},$$
which <u>does not imply</u> ($\approx$ is incoherent with multiplication)
$$u.(u_t + u.u_x) \approx 0.$$
The mistake was to state (1) with equality in **G** (unfortunately suggested by the symbol = in (1)). In the **G**-context one has to be careful in the statement of equations: = or $\approx$: both extend the classical equality.

**\*Charged black holes moving at light speed.** For charged moving at lightspeed black holes physicists have noticed the rather unexpected result that the electromagnetic field vanishes but its energy-momentum tensor does not. This "absurdity" should be clarified: has it its origin in a mathematical mistake or does it point out a breakdown of physics? Rigorous calculations in the **G**-setting [St] show that the field looks like $\sqrt{\delta}$, the square root of a Dirac delta generalized function, while its energy momentum tensor (involving the square of the field) is $\delta$-like. Since $\sqrt{\delta} \approx 0$ the field is infinitesimal but nonzero, thus permitting its energy momentum tensor to be nonvanishing. Then the physically unsatisfactory situation of a vanishing field with nonzero energy momentum tensor is mathematically perfectly clarified by the **G**-setting: the paradox was due to a lack of rigor in the mathematical calculations, again due to a confusion between = and $\approx$: if $G \in \mathbf{G}(\Omega)$ $G \approx 0$ <u>does not always imply</u> $G^2 \approx 0$. For other applications to general relativity see [G-K-O-S, S-V] and references there: impulsive gravitational waves, Schwarzschild and Kerr spacetimes, ultrarelastivistic black holes, geodesics for irregular spacetimes, cosmic strings, "generalized hyperbolicity", "physically reasonable" and "true gravitational" singularities.

**\*a prey-predator model.** The system
    $(\partial_t + \partial_x)u_1 = u_1.u_2$

$$(\partial_t - \partial_x)u_2 = -u_1.u_2$$
$$u_1(x,0)=a(x),\ u_2(x,0)=b(x),$$

describes a population of predators and a population of preys migrating at constant speeds +1 and -1 [C-O,O4]. Using an explicit formula for the solution, which makes sense in the differential algebra $\mathbf{G}(\mathrm{IR}^2)$ even if a and b are irregular objects such as Dirac delta functions modelling populations concentrated at a point, one obtains the following nice results [C-O,O4] in which the equations and the initial conditions are stated with the equality = in $\mathbf{G}(\mathrm{IR}^2)$ and $\mathbf{G}(\mathrm{IR})$ respectively:

Theorem. If a,b $\in \mathbf{G}(\mathrm{IR})$ are positive generalized functions satisfying some natural $\mathbf{L}^1_{loc}$ property then:

- there exists a unique solution in $\mathbf{G}(\mathrm{IR}^2)$ of $\mathbf{L}^1_{loc}$ type;
- this solution is associated to the classical solution when it exists;
- explicit calculations which do not make sense within distribution theory may give "concrete" results when they model realistic physical situations.

For instance let $a(x)=\alpha_1\delta_1(x+1)$ and $b(x)=\alpha_2\delta_2(x-1)$, $\alpha_1,\alpha_2>0$, i.e. at t=0 predators are concentrated at x=-1 and preys are concentrated at x=1, and both populations are modelled with the Dirac delta functions $\delta_1$ and $\delta_2$ (possibly different in the $\mathbf{G}$-context); they meet at t=1, and then move away. Calculations in the $\mathbf{G}$-context show the population of preys has decreased from $\alpha_2$ to $\beta$, $\beta=-2\log\{1-\exp(-\alpha_1/2)+\exp(-(\alpha_1+\alpha_2)/2)\}$ thus $0<\beta<\alpha_2$, and the population of predators has increased from $\alpha_1$ to $\alpha_1+\alpha_2-\beta$, see [C-O,O4]. Note the interesting fact this result does not depend on the "microscopic shapes" of the populations since $\delta_1$ and $\delta_2$ may be different.

This example is typical of a variety of (linear and) nonlinear PDEs for which very nice results of existence-uniqueness can be obtained using the equality = in $\mathbf{G}$, see [O6 and references there] and [Bia,C-O,H-O,K1,K2,K-K-S,O1,O2,O4,O5,O-R,O-W]. A topology "sharp topology" [N-P-S,Bia,Sc] complements these results by adequate stability statements. New "concrete" results can be obtained in physically sensitive cases. In cases classical solutions are known to exist they are recovered with the association.

**3-An ill-posed nonsmooth nonlinear problem resolved by a more precise statement of the equations on physical ground.**

*****Collisions of solids with strong deformation.** This section is issued from the problem of design of armour of battle tanks from numerical simulations of collisions based on the following system of PDEs (given here in 2 space dimensions, see [LR,C3] for 3D). Unknown functions: $\rho$ =mass per unit volume, $(u,v)$=velocity vector, $p$ = pressure, $e$ =total energy per unit volume, ($s_{ij}$)= stress tensor:

1. $\rho_t + (\rho u)_x + (\rho v)_y = 0$
2. $(\rho u)_t + (\rho u^2)_x + (\rho uv)_y + (p - s_{11})_x - (s_{12})_y = 0$
3. $(\rho v)_t + (\rho uv)_x + (\rho v^2)_y + (p - s_{22})_y - (s_{12})_x = 0$
4. $(\rho e)_t + (\rho eu)_x + (\rho ev)_y + [(p - s_{11})u]_x + [(p - s_{22})v]_y - (s_{12}v)_x - (s_{12}u)_y = 0$
5. $(s_{11})_t + u.(s_{11})_x + v.(s_{11})_y = 4/3.\mu(--).u_x - 2/3.\mu(--).v_y + (u_y - v_x).s_{12}$
6. $(s_{22})_t + u.(s_{22})_x + v.(s_{22})_y = -2/3.\mu(--).u_x + 4/3.\mu(--).v_y - (u_y - v_x).s_{12}$
7. $(s_{12})_t + u.(s_{12})_x + v.(s_{12})_y = \mu(--).v_x + \mu(--).u_y - 1/2.(u_y - v_x).(s_{11} - s_{22})$
8. $p = \Phi(\rho, e - u^2/2)$,

where $\mu$ (--)is a function of the variables $s_{ij}$ ($\mu$ is the "shear modulus": constant in the elastic stage, null in the plastic stage) and $\Phi$ is a function. Both functions $\mu$ and $\Phi$ are obtained from experiments on the solid under consideration. All terms in equations. 5,6,7 except the t-derivatives involve products of an Heaviside function and a Dirac delta function which do not make sense in distribution theory and so require the **G**-context. This 2D system is treated by "dimensional splitting" (that is one computes successively in x and y direction) giving rise to 2 systems of 1D equations that still involve the same products. They require the **G**-context in order to compute explicit solutions of the Cauchy problems at the junction of meshes (these particular Cauchy problems are called "Riemann problems") that serve as building blocks to build a numerical scheme of the Godunov type, needed for its efficiency.

**\*Statement of the laws of physics in the G-context**. For the need of the exposition consider as a simplified 1D-model the system of 3 equations classically stated as

(4) $\quad \rho_t + (\rho u)_x = 0, \quad (\rho u)_t + (\rho u^2)_x = \tau_x, \quad \tau_t + u\tau_x = u_x,$

where $\rho = \rho(x,t)$=mass per unit volume, $u=u(x,t)$=velocity, $\tau=\tau(x,t)$=stress. The first equation is the equation of mass conservation, the second one is the equation of momentum conservation, and the third one is a state law of elastic solids in fast deformation (with coefficient chosen equal to 1). Because of the presence of the term $u\tau_x$ in the third equation there appears a product of the kind $H.\delta$ when one seeks a solution of the Riemann problem i.e. the particular case of the Cauchy problem when the given initial data $x \rightarrow$ ($\rho(x,0), u(x,0), \tau(x,0)$) is constant on both sides of a discontinuity. The product $H.\delta$ does not make sense within distributions, so one therefore works in the **G**-setting: we know now that the classical equality has to be replaced by = in **G** or by $\approx$ . Further, as already noted, in the **G**-setting one has to consider an infinity of Heaviside like objects: the formula

(5) $\quad w(x,t) = w_l + (w_r - w_l) H_w (x-ct)$

(where $H_w$ is some Heaviside generalized function depending on w), expresses the fact that the physical variable w takes the value $w_l$ if x<ct and the value $w_r$ if x>ct (discontinuity travelling at constant speed c). From the above we see that the product $u\tau_x$ involves the product $H_u.(H_\tau)'$ that can take different values depending on $H_u$ and $H_\tau$. The product $u\tau_x$ is ambiguous in the **G**-context, in absence of further information: according to its value one obtains different formulas for the solution of the Riemann problem [C3,C4,C5,C-LR,C-LR-N-P,O3]. If all equations in (4) are stated with association then $H_u$ and $H_\tau$ are independent and therefore the result of the insertion of (5) (for $w=\rho, u, \tau$) into (4) is ambiguous: the Riemann problem for the system

(6) $\quad \rho_t + (\rho u)_x \approx 0, \quad (\rho u)_t + (\rho u^2)_x \approx \tau_x, \quad \tau_t + u\tau_x \approx u_x$

is an ill-posed problem backward and forward since there are infinitely many jump formulas. On the other hand the statement of the system with three equalities = in **G** gives non-existence of solutions of the form (5), as this has been noted in the first example in section 2. How to obtain existence and uniqueness at the same time? Let us notice that $H_u$ and $H_\tau$ can be related (therefore fixing the value of the product $H_u.(H_\tau)'$ ) if u and $\tau$ are solutions of (4) when two equations in (4) are stated with = in **G**.

Physicists have observed that shock waves have an "infinitesimal width" of the order of magnitude of a few hundred crystalline meshes. One can therefore isolate (by thought) small volumes inside this width where conservation laws apply. This suggests to state them with (algebraic) = in **G**. On the other hand the state law (3rd equation) has been checked only on a material at rest, i.e. on both sides of the shock wave: by analogy with $H^2 \approx H$ this suggests that one should state it with the weak equality $\approx$ in **G**. Thus it is natural to state in **G** the system of 3 equations as:

$$(7) \quad \rho_t + (\rho u)_x = 0, \quad (\rho u)_t + (\rho u^2)_x = \tau_x, \quad \tau_t + u\tau_x \approx u_x.$$

With this formulation of the problem one obtains the desired existence-uniqueness result and explicit formulas [C3 p63,72, C5]. This method has been used in [LR] for numerical simulations of collisions: a number of numerical solutions are depicted in [Bia,C3,LR]. For contact discontinuities (i.e. mere contacts of a physical nature very different from that of shock waves) this physical idea suggests to state also the state laws with = in **G** since there is no fast deformation: then again one obtains existence-uniqueness. <u>Therefore it is a more precise statement of the equations (following classical ideas from physics) that has solved the lack of uniqueness in the solution by shock waves for this Riemann problem.</u>

Numerical tests show the above is connected with viscosity techniques: state (4) as

$$(8) \quad \rho_t + (\rho u)_x = \varepsilon_1 \rho_{xx}, \quad (\rho u)_t + (\rho u^2)_x = \tau_x + \varepsilon_2 (\rho u)_{xx}, \quad \tau_t + u\tau_x = u_x + \varepsilon_3 \tau_{xx}$$

where the $\varepsilon_i$'s are all "very small". Then the choice of (7) corresponds to the choice $\varepsilon_3 \gg \varepsilon_1$ and $\varepsilon_3 \gg \varepsilon_2$ as this is natural (this agrees with the intuitive conception that "= in **G** is "infinitely" more exact than $\approx$").

We may notice also that the use of = in **G** for the first 2 equations allows to manipulate them freely like in the case of $C^\infty$ functions: for instance immediate calculations imply $u_t + uu_x = \tau_x/\rho$.

**\*A numerical solution of the above 2D-system** ([LR]). A horizontal projectile of copper coming from the left (speed = 3000m/s) collides with a target (at rest, located on the diagonal) made (from left to right) of one layer of aluminium and one layer of copper. The colors represent the pressure: red=about 100 kbars or more, yellow=about 50 kbars, green= about 25 kbars, blue=about 15 kbars, dark blue= between 5 and 10 kbars, white=less than 5 kbars. The pictures are given at 5 and 15 microseconds after the beginning of collision. The process of deformation of the target (whose knowledge is crucial for engineers who design armour) cannot be observed from experiments, which imposes numerical simulations. Further, engineers need to observe the phenomenon during several hundred microseconds under complicated 3D geometries, which imposes very high quality numerical schemes.

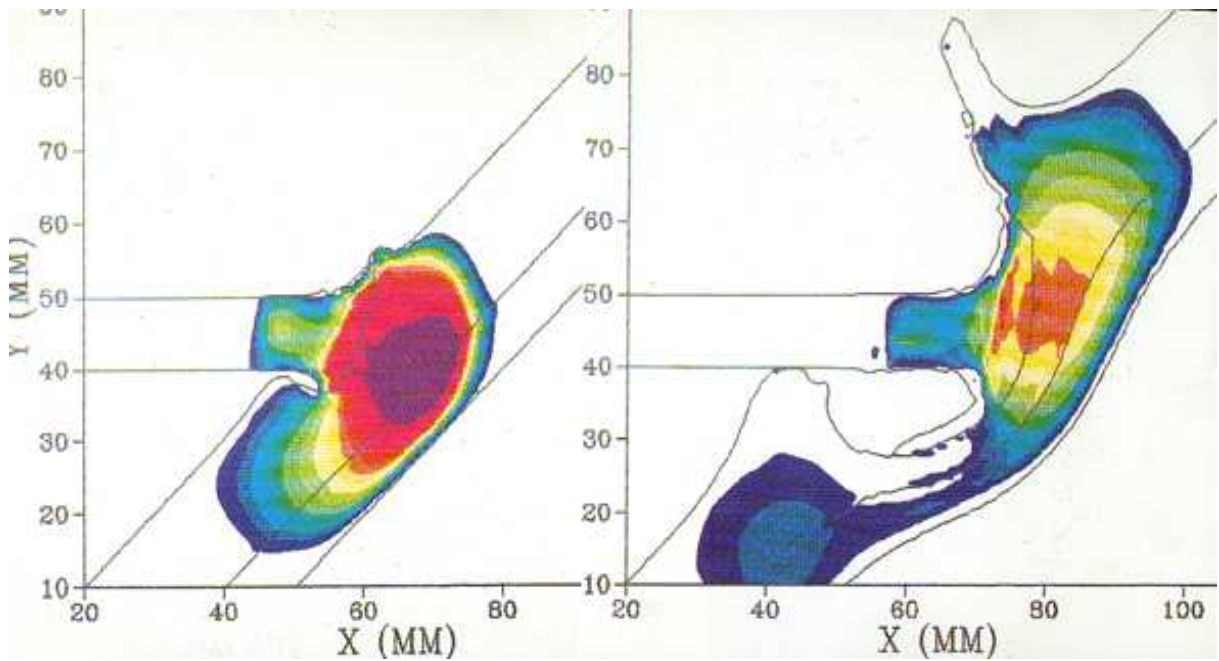

These results are obtained from techniques of numerical engineering that reduce the discretization of 2D-systems to the discretization of simple 1D-systems such as (4) above [LR]. At first the method of "dimensional splitting" consists in treating successively on each cell the x and y directions, thus reducing 2D-systems to two 1D-systems. When the solution of the Riemann problem for a 1D-system cannot be calculated the 1D-system is split into simpler systems by the method of "splitting of equations". The resulting 2D-numerical scheme is validated by comparison with known solutions in particular cases and with experimental data.

**4-Intriguing ill posed–inverse problems depending on a more precise statement of the initial condition.**

**\*A mysterious disappearance (and reappearance).** Consider the nonlinear parabolic equation ($\Delta = \sum_{i=1}^{N} \frac{\partial^2}{\partial x_i^2}$):

$$u_t - \Delta u + u^3 = 0$$

on the strip $\Omega \times ]0, +\infty[$, $\Omega$ open set in $\mathrm{IR}^N$, completed by the initial condition $u(.,0) = f$, f any generalized function in $\mathbf{G}(\Omega)$, with compact support in $\Omega$, and by the boundary condition $u(x,t)=0$ on $\partial\Omega \times [0, +\infty[$, where $\partial\Omega$ is the boundary of $\Omega$. The initial and boundary conditions make sense in the **G**-setting: one seeks a solution in $\mathbf{G}(\overline{\Omega} \times [0, +\infty[)$ (one can define easily nonlinear generalized functions on closed sets). One obtains a nice result [C-L]:

Theorem:
- there exists a unique solution in $\mathbf{G}(\overline{\Omega} \times [0, +\infty[)$;
- this solution is associated to the classical solution when it exists.

Now consider the particular case $u(x,0) = \delta(x-\omega)$, $\omega \in \Omega$; it is proved in [B-F] that in this case, in our language, the solution u is associated to 0 in $\mathbf{G}(\Omega \times ]0, +\infty[)$ and also that for any $t_0 \in ]0, +\infty[$, the restriction $u(.,t_0) \in \mathbf{G}(\Omega)$ is associated to 0 in $\mathbf{G}(\Omega)$: in classical mathematics in which a generalized function associated to 0 is considered as null the effect of the initial condition has disappeared and one concludes to the impossibility of a solution!

Following this particular solution $u \in \mathbf{G}(\overline{\Omega} \times [0, +\infty[)$, the backward problem with initial condition $u(.,t_0)$, $t_0 > 0$, which is associated to 0 in $\mathbf{G}(\Omega)$, produces a solution u such that $u(.,0) = \delta(.-\omega)$ in $\mathbf{G}(\Omega)$: an associated to 0 initial condition at $t = t_0$ has produced a "delta-peak" solution at $t = 0$ for the backward problem !: intuitively $u(.,t)$ looks like some microscopic foam on $\Omega$ for t decreasing from $t_0$ to 0, which "suddenly" condensates at t=0 to produce a particle at the point $\omega \in \Omega$! It is unknown if there is existence-uniqueness in the **G**-context for this backward problem and how to recover the microscopic information on the delta peak hidden in $u(.,t_0) \approx 0$. The absence of a physical origin is a serious drawback since it deprives of the help of physical intuition. See [D,L] for more general nonlinear heat equations. For a better understanding of this phenomenon we consider below explicit solutions only available in case of linear equations.

**\*An interpretation of the classical example of ill-posedness of the inverse problem for the heat equation**. Consider the classical example showing the ill-posedness for the heat equation in the strip $0 < x < \pi$, $t<0$:

$$u_n(x,t) = (1/n) \cdot \exp(-n^2 t) \cdot \sin(nx).$$

This example can be interpreted [C-D] as yielding a generalized function u such that $u(.,0)$ is associated to 0 in $\mathbf{G}([0,\pi])$, and for $t<0$ $u(.,t)$ is not associated to 0. One is in the same situation as above for the nonlinear heat equation, which therefore appears as an interpretation of ill-posedness in the **G**-setting. The association to 0 of u(.,0) hides its genuine "microscopic" structure, which is responsible of the (classical) blow up for t<0 (that appears in the **G**-context in form of a very irregular generalized function).

Consider the backward heat equation
$$(BH) \quad u_t = -k.u_{xx}, \; k>0$$
on the strip $0<x<\pi$, t>0, taking boundary values = 0 on the vertical sides and u(.,0)=f , where f is a function or distribution with compact support in $]0,\pi[$. The classical method of separation of variables and Fourier series puts in evidence an explicit solution u $\in$ **G**$([0,\pi]\times[0,+\infty[)$ of (BH) null on the vertical sides and whose initial condition is associated to f, with a strong lack of uniqueness ( presumably due to the statement of the initial condition in the association sense: is there uniqueness when the initial condition is stated with = in **G**(?) [C-D]. The Laplace equation $u_{tt} = -u_{xx}$ presents a similar mathematical picture both forward and backward (Hadamard counterexample of ill posedness).

**5. Objects recognition from acoustic or electro-magnetic waves.** This topic (motivated by submarine and aerial detection) has motivated a number of theoretical and numerical works in the **G**-setting on linear systems of PDEs with discontinuous coefficients (modelling the object as an irregularity in the medium): existence, uniqueness, propagation of singularities and microlocal analysis, numerical solutions and comparison with experimental data. Up to now this concerned linear equations not in the scope of this talk: the "classically forbidden multiplications of distributions" arise in the products of the coefficients and the derivatives of the solution that both are singular on the discontinuities of the medium. See a system of linear acoustics in [C3 p 28] and numerous articles of mathematics [O4 §17,N-P-S,O6 and arXiv for recent articles].

**6. Final comments.**
*In "one word" what to retain on these nonlinear generalized functions?* Consider the significant example of the "Dirac delta function" located at the origin: in the **G-**context there is an infinity of them; the **G-**context permits the idealization needed for the sake of simplicity (to make mathematical calculations possible) at the same time as it retains some internal structure making the difference between them (which is absent in distribution theory where there is only one Dirac delta distribution). This refined description is at the basis of most applications (and of the fact this theory escapes from the Schwartz impossibility result); this refinement disappears when generalized functions are considered modulo the association: from the various Dirac delta generalized functions one recovers only one "classical Dirac delta like object", without internal structure.

In the definition of the full algebra one considers any distribution as having "all possible internal structures", which amounts to "no internal structure at all" and permits the canonical inclusion $\mathbf{D'}(\Omega) \subset \mathbf{G}(\Omega)$, without a change in the nature of the concept of distribution. Concerning the simplified or special algebra $\mathbf{G}_S(\Omega)$ a non canonical choice of internal structure for the distributions has to be done in order to obtain an embedding $\mathbf{D'}(\Omega) \subset \mathbf{G}_S(\Omega)$, but one gains in simplicity and this is perfect when only distribution like objects, defined modulo the association, are needed.

***The G-theory as a smoothing method**. Its mathematical originality is that it displays very simple differential-algebraic properties: $\mathbf{G}(\Omega)$ is a differential algebra in which one may compute like in the familiar case of $\mathbf{C}^\infty$ functions. This is achieved by use of adequate asymptotic estimates with respect to a regularization parameter $\varepsilon \to 0^+$ (in the simplified $\mathbf{G}_S$-version; the full **G**-version uses a more technical parameter).The need to eliminate ill-posedness in calculations involving products of distributions forces to retain some information on the internal structure of the mathematical objects under consideration. To this end the definition of these generalized functions is not a limit process as usual, but a quotient for a suitable equivalence relation.

**\*Connections with physics**. This new kind of "microscopic information" can often be easily obtained from physics in that it can be implicit in the basic usual physical concepts (at least in classical mechanics [C3 and articles] and general relativity[S-V and articles]).

An idea is that the genuine physical objects should be represented by a very small unknown value of the parameter $\varepsilon$ (that has to be arbitrarily small in the definition of **G**($\Omega$)) to obtain a needed idealization (both by ignorance and for the need of simplicity). One may have in mind that intuitively we work for fixed, unknown, very small, $\varepsilon >0$, and then let $\varepsilon \to 0^+$ only at the very end of the calculations to get the final numerical results.

**\*Conclusion**. This theory of nonlinear generalized functions is a mathematical tool more refined than classical functions or distributions; it provides a deeper description of the physical reality. This permits in some cases to resolve ambiguities that appear from a use (often without mathematical rigor) of classical mathematics out of its domain of relevance first by pointing out the nature the lacking information, and then by helping to choose it. The strategy for this choice consists often in going back as far as the physical origin of the problem in order to discover its precise mathematical statement in the **G**-context.

**7. Appendix. An attempt to explain the basic idea by a very simple intuitive analogy.**

Consider an extremely small (may be unknown) positive physical quantity $\varepsilon >0$ and the quantities noted "$\infty$":=$1/\varepsilon$, "$0_1$":=$\varepsilon$, "$0_2$":=$2\varepsilon$, "$0_3$":=$\sqrt{\varepsilon}$, "$0_4$":=$\varepsilon^2$. Now imagine one needs an abstract mathematical idealization of these quantities (such as the one consisting in modelling an extended object concentrated in a very small region by a Dirac delta function). Such idealizations are needed to apply evolved mathematics (if the above quantities are replaced by complicated functions and if one seeks solutions of equations in which they are involved; of course no idealization is needed to do products of these objects, which is only presented here as a trivial analogy).

The analog of the idealization provided by classical analysis would be to assign the abstract value 0 to all the "$0_i$" 's (since $\varepsilon$ is extremely small) like the classical idealization in the support of a Dirac delta distribution (= a point without any "structure"). Then in this context the product $0 \times \infty$ does not make sense (4 different "physical" results would be expected according to the index $i$ !). It is clear this is due to an abusive idealization that has destroyed too much information.

The analog of the idealization provided by the **G**-context would be to state all the "$0_i$" 's different (by retaining some information on the way they converge to 0 as $\varepsilon \to 0$), and all associated to 0. Then the products "$0_i$"$\times \infty$ can be different, and coherence with classical analysis is obtained via the association. This is correct provided one checks that the association and the equality of classical analysis reflect the same degree of idealization.

Such a mathematical context can be defined as follows: set

E ={f: $\varepsilon \to$ f($\varepsilon$) such that $\exists \eta >0, C >0, N \in$ IN such that | f($\varepsilon$)| $\leq$ C/($\varepsilon^N$) if $0< \varepsilon <\eta$ },

N ={f$\in$ E such that $\forall q \in$ IN $\exists \eta >0, C >0$ such that | f($\varepsilon$)| $\leq$ C.$\varepsilon^q$ if $0< \varepsilon <\eta$ }.

N is an ideal of the algebra E and it suffices to consider as mathematical context the quotient algebra E/N. An element (f+N) of E/N is said to be "associated to 0" iff f($\varepsilon$)$\to$0 as $\varepsilon \to 0$ (notice the great loss of information relatively to N, i.e. relatively to = 0 in E/N). A more complicated but basically similar construction gives the simplest variants $\mathbf{G}_S(\Omega)$, see [C1].

jf.colombeau@wanadoo.fr; 33 rue de la Noyera, pavillon 17, 38090, Villefontaine, France.

**References.**

[Bia]H.A.Biagioni. A Nonlinear Theory of Generalized Functions. Lecture Notes in Math 1421, Springer Verlag, 1990.

[B-F]H.Brezis,A.Friedman. Nonlinear parabolic equations involving measures as initial conditions .J Math.Pures Appl.62,1983,p73-97.

[C1]J.F.Colombeau. Generalized functions and infinitesimals, arXiv 2006, FA/0610264.

[C2]J.F.Colombeau. A multiplication of distributions, J.Math.Ana.Appl. 94,1,1983, p96-115. New Generalized Functions and Multiplication of Distributions. North Holland, Amsterdam, 1984. Elementary Introduction to New Generalized Functions. North Holland, Amsterdam, 1985.

[C3]J.F.Colombeau. Multiplication of Distributions. Lecture Notes in Maths 1532. Springer Verlag 1992.

[C4]J.F.Colombeau. Multiplication of distributions. Bull. AMS. 23,2, 1990, p251-268.

[C5]J.F.Colombeau. The elastoplastic shock problem as an example of the resolution of ambiguities in the multiplication of distributions J. Math Phys 30, 90, 1989, p2273-2279.

[C-D] J.F.Colombeau-A.Delcroix. private communications.

[C-L]J.F.Colombeau,M.Langlais. Generalized solutions of nonlinear parabolic equations with distributions as initial conditions J.Math.Anal.Appl.145,1,1990,p186-196.

[C-LR]J.F.Colombeau,A.Y.LeRoux. Multiplication of distributions in elasticity and hydrodynamics. J.Math.Phys. 29,2,1988,p315-319.

[C-LR-N-P]J.F.Colombeau,A.Y.LeRoux,A.Noussaïr,B.Perrot. Microscopic profiles of shock waves and ambiguities in multiplication of distributions. SIAM J. Num.Anal. 26,4,1989,p871-883.

[C-O]J.F.Colombeau,M.Oberguggenberger. Hyperbolic system with a compatible quadratic term, delta waves and multiplication of distributions. Comm.in PDEs 15,7,1990, p905-938.

[D]H.Deguchi. Generalized solutions of semilinear parabolic equations.Monatshefte für Math. 146,p279-294,2005.

[G-K-O-S]M.Grosser,M.Kunzinger,M.Oberguggenberger,R.Steinbauer. Geometric Theory of Generalized Functions with Applications to General Relativity. Kluwer, Dordrecht-Boston-New York, 2001.

[H-O]R.Hermann,M.Oberguggenberger. Ordinary differential equations and generalized functions.in Grosser,Hörmann,Kunzinger (eds). Nonlinear Theory of Generalized Functions. Chapman § Hall. Research Notes in Math; 401,1999,p85-98.

[Ho]R.F.Hoskins, J.Sousa-Pinto. Distributions, Ultradistributions and Other Generalised Functions. Ellis-Horwood, New York-London, 1994.

[K1]I.Kmit. Initial boundary value problems for semilinear hyperbolic systems with singular coefficients. arXiv 2004.

[K2]I.Kmit. Generalized solutions to hyperbolic systems with nonlinear conditions and strongly singular data. Integral Transforms Spec. Funct. 17,p177-183, 2006.

[K-K-S]I.Kmit,M.Kunzinger,R.Steinbauer. Generalized solutions of the Vlasov-Poisson system with singular data. ArXiv Jan.2006.

[La]M.Langlais.Generalized functions solutions of monotone and semi-linear parabolic equations .Monatshefte fûr Math.,110,1990, p117-136.

[LR]A.Y.LeRoux. Phd thesises held at the University of Bordeaux 1, to be found from the university library: Cauret 1986, Adamczewski 1986, De Luca 1989, Arnaud 1990, Baraille 1992.

[N-P-S]M.Nedjelkov,S.Pilipovic,D.Scarpalezos. The linear theory of Colombeau generalized functions Pitman Research Notes in Math. 385,Longman,1998.

[O1] M.Oberguggenberger. Generalized solutions to semilinear hyperbolic systems. Monatshefte fûr Math. 103, 1987, p133-144.
[O2] M.Oberguggenberger. The Carleman system with positive measures as initial data; generalized solutions. Transport Theory and Statistical Physics, 20, 1991, p177-179.
[O3] M.Oberguggenberger. Case study of a nonlinear, nonconservative, nonstrictly hyperbolic system. J. of Nonlinear Analysis, T.M.A., 19, 1992, p53-79.
[O4] M.Oberguggenberger. Multiplication of Distributions and Applications to PDEs. Pitman Research Notes in Math 259, 1992, Longman.
[O5] M.Oberguggenberger. Generalized functions, nonlinear PDEs and Lie groups. In RS.Pathak (ed), Geometry, Analysis and Applications. World Sci. Pub. Singapore, 2001, p271-281.
[O6] M.Oberguggenberger. Colombeau solutions to nonlinear wave equations. arXiv 2006, AP/0612445.
[O-R] M.Oberguggenberger, F.Russo. Nonlinear stochastic wave equations. Integral Transforms and Special Functions 6, 1998, p71-83.
[O-W] M.Oberguggenberger, Y.G.Wang. Delta waves for semilinear hyperbolic Cauchy problems. Math. Nachr. 166, 1994, p317-327.
[P] I.Petrovsky. Lectures on PDEs, Interscience Pub, (first english edition 1954).
[Sc] D.Scarpalezos. Colombeau's generalized functions: topological structures, microlocal properties, a simplified point of view I. Bull.Cl.Sci.Math.Nat. 25, p89-114, 2000. Same title II. Pub.Instit.Math.Beograd. 76, p111-125, 2004.
[St] R.Steinbauer. The ultrarelastivistic Reissner-Nordström field in the Colombeau algebra. J. Math. Phys. 38, p1614-1622, 1997. ArXiv.
[S-V] R.Steinbauer, JA.Vickers. The use of generalized functions and distributions in general relativity. Class.Quant.Grav.23, pR91-114, 2006, and numerous references there, many in ArXiv. .